\documentclass[12pt]{article}
\usepackage{graphics}
\input epsf
\usepackage{amsmath,color,fancybox,graphics,graphpap,rotating}
\definecolor{yellow}{rgb}{0.95,0.75,0.1}
\definecolor{red}{rgb}{1,0,0}
\definecolor{darkred}{rgb}{0.5,0,0}
\definecolor{green}{rgb}{0,1,0}
\definecolor{blue}{rgb}{0,0.5,1}
\definecolor{bgcolor}{rgb}{0.9,0.9,0.999}

\newcommand{\be}{\begin{eqnarray}}
\newcommand{\ben}{\begin{eqnarray}\nonumber}
\newcommand{\ee}{\end{eqnarray}}
\newcommand{\nee}{\nonumber \end{eqnarray}}

\begin{document}
\title{
\begin{flushright} 
UAHEP044
\end{flushright}
\vskip 1cm
A Susy Phase Transition as Central Engine\footnote{Extended version of a 
talk presented at the conference on Fundamental Physics from Clusters of Galaxies,
Fermilab, Dec 9-11,2004}}
\author{L. Clavelli\footnote{lclavell@bama.ua.edu}\\
Department of Physics and Astronomy\\
University of Alabama\\
Tuscaloosa AL 35487\\ }
\maketitle
\begin{abstract}
   For several decades the energy source powering supernovae and gamma ray bursts has been a troubling mystery.  
Many articles on
these phenomena have been content to model the consequences
of an unknown ``central
engine'' depositing a large amount of energy in a small region.
In the case of supernovae this is somewhat
unsettling since the type 1a supernovae are assumed to be ``standardizable
candles'' from which important information concerning the dark energy
can be derived.  It should be expected that a more detailed understanding
of supernovae dynamics could lead to a reduction of the errors in this
relationship. Similarly, the current state of the standard model theory of gamma ray bursts, which in some
cases have been associated with supernovae, has conceptual
gaps not only in the central engine but also in the mechanism for jet collimation and the lack of baryon
loading.  We discuss here the Supersymmetric (susy) phase transition 
model for the central engine.
\end{abstract}
\renewcommand{\theequation}{\thesection.\arabic{equation}}
\renewcommand{\thesection}{\arabic{section}}
\section{\bf Introduction}
\setcounter{equation}{0}
  
    The ``central engine'' puzzle of violent astrophysical events 
has been evident
\cite{Bethe} for decades.  For a period, it was assumed that
the neutrinos released from nuclear fusion could provide the
energy needed to blow off the stellar mantle in a supernova
but detailed monte-carlos never succeeded in demonstrating
the required explosion.  The neutrinos were too few and too
weakly interacting.  More recently, strong magnetic fields
of unknown origin have been discussed as a possible part of 
the solution \cite{Duan}.  The situation is such that 
new physics input may be needed.
One is reminded of the nineteenth century mystery 
of the sun's ``central engine'' and of the long ignored
mid-twentieth century prediction of neutron stars. 
In the last few decades new physics proposals for the 
enormous energy release in gamma ray bursts such as, 
for example, the quark star model of 
ref.\cite{Sannino} have, perhaps, not
received the full discussion that they may deserve. 

   In recent years it has become increasingly likely that the expansion of 
the universe is accelerating in a way consistent with an
 interpretation in terms of a positive vacuum energy density of approximate
magnitude
\be
         \epsilon = 3560 MeV/m^3 .
\label{vacenergy}
\ee
A natural value that might have been expected for this quantity would be
some 124 orders of magnitude greater:
\be
      M_{Planck}^4 = 10^{127} MeV/m^3 .
\ee

\begin{figure}[ht]
\begin{center}
\epsfxsize= 4.5in 
\leavevmode
\epsfbox{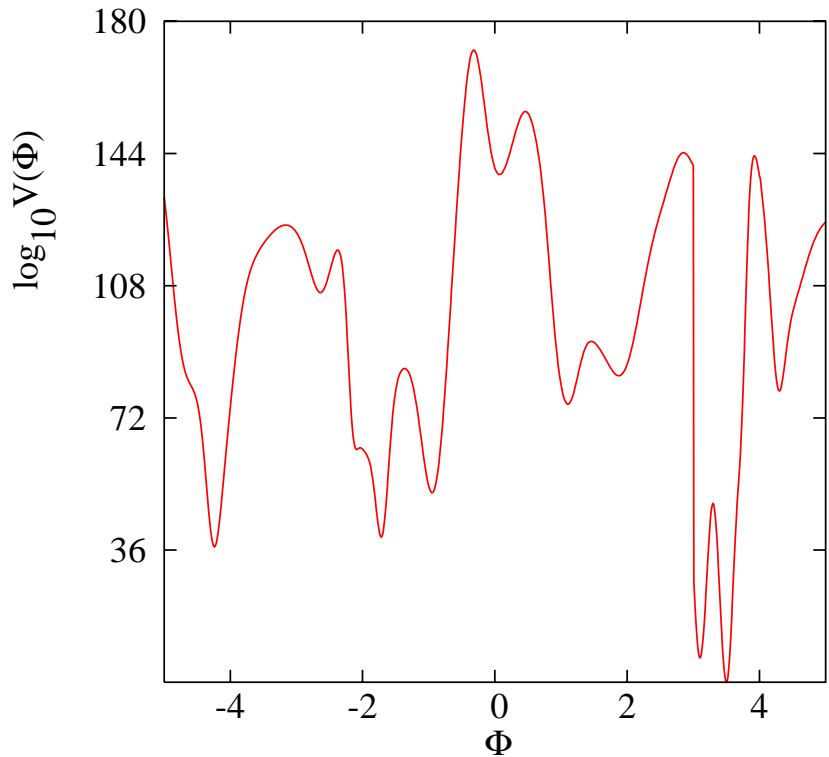}
\end{center}
\caption{A schematic representation of the effective potential in the string landscape
picture.  The potential is measured in units of MeV/m$^3$ and
the y axis has a broken scale taken to be linear in V at low values of the potential.
The exact susy phase has zero vacuum energy.}
\label{landscape}
\end{figure}

A popular attitude towards this huge mismatch is the string landscape picture
in which the effective potential as a function of the myriad scalar fields
of string theory has a huge number of local minima most of which are of order 
of $M_{Planck}^4$.  However, about one in $10^{124}$ of these minima would be
expected to have vacuum energy density comparable to $\epsilon$.  
If the vacuum energy were much larger than eq.\ref{vacenergy} 
the universe would be torn apart by the large acceleration\cite{Weinberg}
before galaxies and life would have had time to evolve.
  If there were
such a huge number of universes, either temporally or spatially separated,
it would not be surprising that we find ourselves in one with at most a
mild acceleration.  We could never be conscious of any other.

Furthermore, if all the parameters of the effective potential were
dynamically determined as in superstring theory, one would expect quantum 
transitions between the various minima.  Such transitions
between string vacua with differing amounts of supersymmetry
have already received some attention \cite{Kachru}.
The basic string theories and, most likely, those with the
 absolute minimum of the
effective potential have exactly supersymmetric (susy) vacua with vanishing
vacuum energy.  Unfortunately, these vacua are also inhospitable to life.

    A key difference between a supersymmetric universe and our broken
susy world is the reduced importance of the Pauli principle in the former.
This is due to the fact that, in a susy universe, whenever fermions are 
forced into higher energy levels, it becomes
energetically favorable for them to convert in pairs into scalars which can drop 
into the ground state in arbitrary numbers.  Thus heavy nuclei and atoms 
would consist mostly of scalar nucleons and selectrons.  Without the need to
put additional particles into ever higher energy levels, fusion would occur at a
greatly increased rate thus greatly reducing the lifetime of main sequence stars
probably below the time required for planets to form and life to evolve.
Furthermore, most nuclei that are stable with fermionic constituents would be highly
beta unstable in a susy vacuum due to the absence of an effective Pauli 
principle.  In addition scalar electrons would all drop into ground state 
s wave states where
they would have an increased probability of being K captured. Thus an exactly
supersymmetric universe would be an impossibly toxic environment for human
evolution.

\vspace{0.1in}
\begin{figure}[ht]
\begin{center}
\epsfxsize= 4.5in 
\leavevmode
\epsfbox{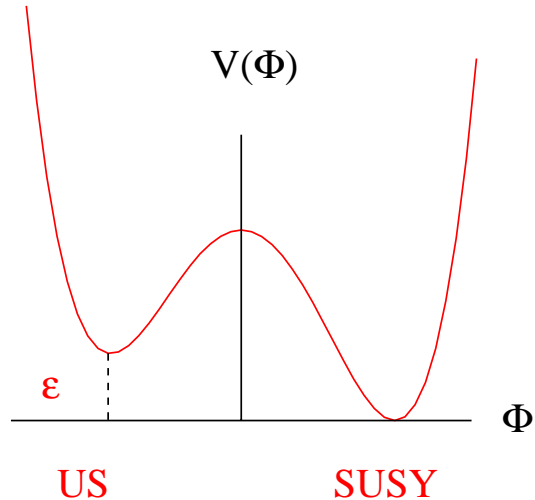}
\end{center}
\caption{Double well potential in vacuo with a susy minimum}
\label{dwellinvac}
\end{figure}

    Evidently, because of statistics, luck, or some other reason, there exists
a local minimum which is not supersymmetric but also does not have too high
a vacuum energy.  
We do not consider here the possibility
that there might be other local minima with negative vacuum
energy leading to other unpleasant vacuum decays.

    Such an effective
potential, the so called string landscape \cite{Susskind}, can be schematically 
represented as in figure \ref{landscape}.  The horizontal axis
represents the space of scalar fields projected onto one dimension.  The absolute
minimum is supersymmetric and has vanishing vacuum energy while a second
local minimum of broken susy and near vanishing vacuum energy is available.  
In such a broken susy world life has evolved on earth.  
This is not to imply that we understand the fine tuning required to make the susy 
breaking scale of $100$ GeV, perhaps in conjunction with another
dark energy component, consistent with the low vacuum energy of 
eq. \ref{vacenergy}.
Furthermore, unlike the 
simplified picture of figure \ref{landscape}, the broken
susy minimum may be separated from the susy minimum by a great distance in $\phi$
and there may be many other intervening local minima.  This would quantitatively
affect the probability per unit time of vacuum decay but would not qualititatively 
change the ultimate outcome.

\begin{figure}
\begin{center}
\epsfxsize= 4.5in 
\leavevmode
\epsfbox{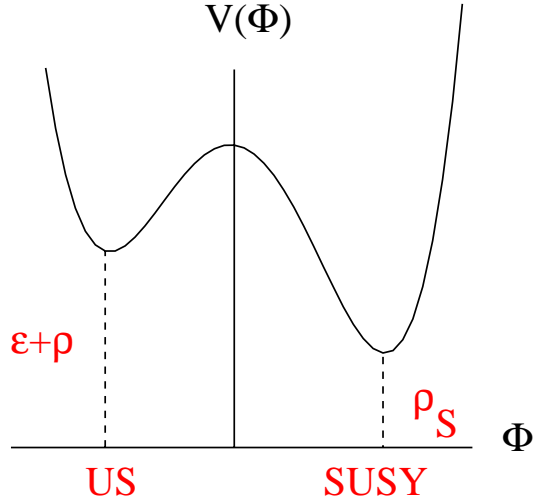}
\end{center}
\caption{Susy double well in the presence of dense matter}
\label{dwellinmat}
\end{figure}
\vspace{0.1in}

     If we zoom in on the region near the potential minimum, it might look in vacuo, like
the double well potential of figure \ref{dwellinvac} treated for the decay of the 
false vacuum by Coleman \cite{Coleman} and others several decades ago.  It has been
shown in lower dimensional theories that the transition rate is enhanced in the
presence of matter \cite{Gorsky} and it has been speculated that the same enhancement 
would occur
in dense matter in any space-time dimension \cite{Voloshin}.  A new ingredient, which may be key
to proving this enhancement and which has not been considered in previous years, is the role 
of the Pauli principle in dense matter.  If a system of fermions in its ground state
has energy density $\rho$, the ground state energy density of the corresponding
susy system with the same baryon and lepton numbers whould be $\rho_s$.   The 
difference between the ground state energy densities in the two phases is the 
excitation energy density of the fermions in the normal phase.  This assumes that
the common mass of particles and sparticles in the exact susy phase is equal to
that of the particles in the broken susy phase.   In superstring theory the ground 
state supermultiplets are massless so it is reasonable that the common mass is
light.  If the common mass were lighter than that of the fermions in the broken susy 
phase the energy release in the phase transition would be increased above that of our
current studies.   If the common mass were
heavier than that of the fermions in the broken susy phase the broken susy phase 
in figure \ref{dwellinmat}
would be the stable minimum and there would be no transition to exact susy.

    In a region of dense matter, the effective
potential may take the form of figure \ref{dwellinmat}.
If one is in the false vacuum of broken susy,
bubbles of true vacuum (exact susy) are constantly being
produced with a steeply falling distribution in bubble radius, r.
At creation, or at any later stage in its development, a bubble of radius r 
will expand or contract depending on which behavior is energetically favorable.
The condition for expansion depends on the surface tension, S, of the bubble
and the energy density in the region immediately outside the bubble.
Consider, for example, a bubble of exact susy in a larger region of broken
susy.  Outside the bubble the energy density will be $\epsilon + \rho$
where $\epsilon$ is the vacuum energy  and $\rho$ is the outside ground state
matter density if any.  If the susy bubble were to make a virtual expansion
into an infinitesimally larger spherical shell, its ground state 
energy density would
be $\rho_s$.  Classically, the ground state energy after such a virtual
expansion minus the previous ground state energy is
\be
    \Delta E = \frac{4 \pi}{3}\left( (r+\delta r)^3 - r^3\right) (\rho_s - \rho - \epsilon)
                 + 4 \pi S \left( (r+\delta r)^2 - r^2 \right)
\ee
or
\be
    \Delta E = - 4 \pi r \delta r \left( r (\epsilon + \Delta \rho) - 2 S \right)  
\label{deltaE}
\ee
where we have put 
\be
    \Delta \rho = \rho - \rho_s
\ee

Classically therefore, the system will find it energetically advantageous to expand if
$r > \frac{2S}{\epsilon + \Delta \rho}$.  Similarly, the bubble will contract if its
radius is below this density dependent critical value.   A more exact instanton calculation \cite{Coleman}
in vacuo ($\rho = \rho_s = 0$) replaces $2S$ by $3S$ in eq.\ref{deltaE}.  One would, therefore, expect the critical radius for a susy bubble to be
\be
    R_c = \frac{3S}{\epsilon + \Delta \rho}
\ee
There is an implicit assumption here that the surface tension, S, is not
significantly density dependent as in the low density limit.  A more careful treatment of the surface tension is in the process of being explored.
In regions of high density, and therefore high $\Delta \rho$, the critical
radius will be much reduced from its vacuum value thus enhancing the
phase transition rate.  At creation or at any later stage, a bubble will  expand if $r>R_c$.
In vacuo or ignoring the effect of the Pauli principle, $\Delta \rho=0$.  In a 
homogeneous region, if a bubble is created at greater than the critical radius, it will expand indefinitely.  If however, the bubble comes to the boundary of a 
dense region outside of which $\rho$ and $\Delta \rho$ are zero, the critical
radius jumps discontinuously to its vacuum value, effectively confining the bubble to the high density region.         
In vacuo, the probability per unit time per unit volume of nucleating a
bubble of true vacuum of critical radius or greater has been given 
\cite{Coleman} in the thin wall approximation as $A e^{-B}$
where $B \approx {R_c}^4 \epsilon$.
If we naively carry this formula over to the case 
of a susy phase transition in a homogeneous body of
volume V, we would have
\vspace{0.2in}
\be\nonumber
   \frac{1}{N} \frac{dN}{dt} = AV e^{-(\frac{\rho_1}{(\epsilon + \Delta \rho)})^3}
\label{AEMB}
\ee

where $\rho_1$ is an undetermined parameter of dimension [$E^4$]. 
We expect the important features of this expression to survive the extension
to a body of non-uniform density with a density dependent surface tension and
a non-negligible wall thickness between the normal and susy phases.  These
important features are that the phase transition rate per unit volume
increases rapidly with $\Delta \rho$ up to some density $\rho_1$ and then
saturates.  At higher $\Delta \rho$, the transition rate is proportional
to the volume of the body.

     In a recent article \cite{CK}, we have proposed that such a phase transition
between our normal phase of broken susy and a phase of exact susy is the
central engine of gamma ray bursts. Many of the zeroth order characteristics
of the bursts can be readily understood in this model.  Our current thinking
is that
$\rho_1$ should be of order of the $\Delta \rho$ that would be expected in a 
white dwarf star.  Then, in less dense matter the phase transition rate
would be exponentially surpressed while in more dense matter such as
neutron stars or ordinary heavy nuclei the phase transition rate would
be greatly suppressed by the volume factor.  In the present state of
the model, the constant A is a free parameter 
fit to the observed rate of gamma ray bursts.  

    The susy phase is a rich new world of phenomenology which has only just
begun to be explored.  Unlike our world which is dominated by the Pauli 
principle, the susy world would have a strikingly
different particle, nuclear, and atomic physics.
Ignoring the nuclear effects, which were discussed in outline form in
\cite{CK}, the key features of the susy phase transition model of grb's
are 
 
\begin{enumerate}
\item {A white dwarf star with a high level of fermion degeneracy
 makes a phase transition to a state of exact supersymmetry where
particles and their susy partners (sparticles) have equal masses.}
\item {Electron pairs undergo quasi elastic scattering to 
 selectron pairs via photino exchange.}
\be
  e^{-} e^{-} \rightarrow {\tilde e}^{-} {\tilde e}^{-}
\label{sigma}
\ee
\item {Uninhibited by the Pauli principle, the selectrons
fall into the ground state emitting photons which can
penetrate into the broken-susy world.}
\item {Since selectrons and photons are Bosons, at least some 
amount of jet structure is produced by stimulated emission.}
\item {With no further support from electron degeneracy, 
The star collapses below the\\
Schwarzschild radius and becomes a black hole.}
\end{enumerate} 

We can briefly elaborate somewhat on each of these items as follows:
\begin{enumerate}
\item{Our current thinking is that $\rho_1$ is such that the
vast majority of susy phase transitions take place in white
dwarf stars.  We have also considered the possibility that
neutron stars are the progenitors of the gamma ray bursts.}
\item{The pair conversion process has been recently calculated 
\cite{CP} in the susy 
phase where electrons and selectrons have the same mass.  
Amplitudes in the normal phase where the electron mass is
negligible compared to the selectron mass were calculated
earlier by Keung and Littenberg \cite{Keung}.  At the same
time in the susy phase, quarks will convert to scalar quarks
via gluino exchange.  A full conversion to scalar nucleons 
will be third order in the strong fine structure constant
although a significant enhancement is to be expected from
low lying quark-squark bound states (pioninos). Since, even
at white dwarf densities, separate nuclei are outside the range of
strong interactions, strong conversion takes
place at first only within nuclei.} 
\item{ The photons can penetrate the bubble wall since they
are light in both phases.  The selectrons, however,
being extremely massive in the broken susy phase are confined
to the interior of the bubble.  The resulting 
enormous energy release due to the relief from Pauli blocking
can re-ignite fusion in the cold
star and provide the energy to accelerate outward any
circumstellar material which may be present.  Thus the susy
phase transition may provide a fundamental explanation for the
``central engine'' powering gamma ray bursts in the
``collapsar'' and ``cannonball'' models.  However, in the
current model, the gamma ray bursts can erupt even in the
absence of significant circumstellar matter.}
\item{Beam collimation due to stimulated emission of bosonic
particles is familiar from laser physics.  The fundamental
enhancement mechanism derives from the behavior of bosonic
creation and annihilation operators discussed below.}
\item{In standard model astrophysics, isolated white dwarfs are totally
stable due to the Pauli principle and there should be no black holes below 
the Chandrasekhar limit of $1.44$ solar masses.  In the current model however, 
once the Pauli blocking
has been lifted, the typical white dwarf of solar mass and earth radius
will collapse to a black hole in (classically) about $1.5$ s.  This free collapse time
is tantalizingly close to the $2$ s dip in the duration distribution of
gamma ray bursts.  We are currently investigating this as well as
other mechanisms for the dip.}

\begin{figure}[htb]
\begin{center}
\epsfxsize= 4.5in 
\leavevmode
\epsfbox{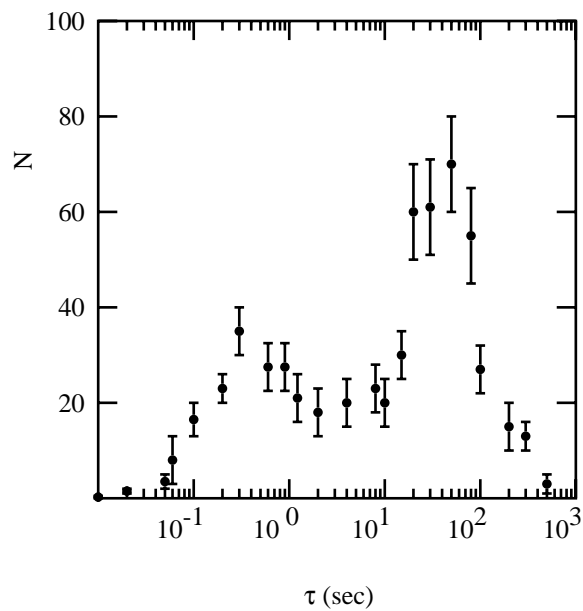}
\end{center}
\caption{duration distribution of gamma ray bursts after ref.\cite{durationdist}}
\label{duration}
\end{figure}

   In conventional astrophysical models for the bursts, the 
duration distribution shown in figure \ref{duration} is sometimes assumed to come from a viewing 
angle dependence although the existence and location of the dip is not easily predicted \cite{Zhang,Yamazaki}.  

\end{enumerate}  

    In ref.\cite{CK} we were able to easily show that the energy release from 
a susy phase transition in a white dwarf star would result in an at least partially
collimated burst of MeV level gamma rays (the average kinetic energy in the degenerate
electron sea) with a total energy from electrons
of about $10^{50}$ MeV (the total kinetic energy in the electron sea).  
The minimum time scale of the event would be about
$0.02$ s, the time for a light ray to traverse the radius of the typical 
white dwarf star (solar mass, earth radius).   The predicted numbers are in 
agreement with observations making the common assumption that the burst has 
about a $5^\circ$ opening angle.  

     One would expect that any new idea for the central engine would be welcomed 
in astrophysics
but it is also fair to critically compare the susy phase transition model with more
conventional models for the gamma ray bursts to see which is more speculative.  
The long duration bursts ($> 2$ s)
have been extensively investigated in terms of the ``collapsar'' model \cite{collapsar}.
In this model one begins with a ``failed supernova'' that has left a large amount
of matter in one or more shells surrounding a 
collapsed core.  The shells fall onto the dense core reigniting fusion 
the energy release from which is transferred to the infalling material causing
them to rebound at relativistic
speeds colliding repeatedly with the core and with each other to produce the observed
gamma rays.  The collapsar model has evolved to incorporate some features of the
``cannonball'' model of Dar and DeRujula \cite{DeRujula}.  In this latter model
a large amount of stellar material is ejected from a star and the gamma rays
are produced externally when this relativistic material collides with circumstellar matter.  
  
      Perhaps conventional astrophysical scenarios can be constructed in which the
zeroth order quantitative observations mentioned above can be fit to reasonably
valued parameters but, in these scenarios, several questions remain to be resolved: 

\begin{enumerate}
\item{What is the ``central engine'' (energy release mechanism)?} 
Newtonian gravity by itself does not provide more energy to
a falling object than its original potential energy.
In Einsteinian gravity we are familiar with black holes continually pulling in surrounding 
matter but is there
a mechanism for a black hole, or incipient black hole, to emit near stellar sized objects 
at relativistic speeds?   
Does conventional fusion provide a sufficient energy release mechanism?
In an ``evolved'' core consisting of the nuclear ash of earlier burning, the remaining nuclei are
relatively heavy and relatively little energy is available even from fusion reignition 
at elevated temperatures.  Iron is the endpoint of conventional fusion and further
fusion reactions absorb rather than release energy.  
\item{What is the collimation mechanism?}
In the cannonball model the collimation is produced by the large Lorentz parameters
of the ejected matter ($\gamma \approx 100 - 1000$).  This would explain a collimated
nature of the gamma ray bursts but what force can outwardly accelerate a macroscopic body
to such speeds?  Perhaps the susy phase transition photons in the presence of
circumstellar material could produce the proposed cannonball acceleration. 
In the collapsar model, if the infalling shells are spherical or
disk-like, it is difficult to construct a monte carlo that would transfer a huge amount
of momentum into any one direction, even if there is a preferred axis such as the
normal to an accretion disk.  Here also, some combination of accretion disk models
with a susy central engine could be productive.
\item{What causes the lack of ``baryon loading''?}
In conventional astrophysical models, even if a large amount of energy is available,
it is difficult to deposit a large fraction of this energy into a limited range of
the gamma ray spectrum as is observed.  In a supernova for example most of the energy
goes initially into kinetic energy of baryons and heavy nuclei and from there into a 
broad range of photons mostly well below the gamma ray region.  In heavy ion colliders
where nuclei are given Lorentz parameters of up to $100$, no events are observed where
the energy is largely deposited into gamma rays.
\item{What are the short duration bursts ($< 2$ s)?}
The collapsar model seems incapable of incorporating the short duration bursts 
suggesting that they may be due to a totally different mechanism.  Recently, 
several ideas \cite{Lee} have been put forward to understand these short bursts.
In one of these it is suggested
that about $10^{50}$ ergs (depending on accretion disk viscosities
and an assumed efficiency of $1 \%$)
could be converted from $\nu \overline{\nu}$ into an $e^+e^-$
plasma which could then be made available for the production of
a relativistic fireball.  This model could be in line with the
``cannonball'' model of Dar and DeRujula \cite{DeRujula} which
involves the acceleration of a relativistic fireball away from a 
progenitor star.  However, it still leaves open some major questions.
The central engine is still unidentified and, even if a huge
electron-positron cloud is available, what force can accelerate it
outward and why don't the electrons and positrons annihilate long before they 
produce gammas from a collision with circumstellar material?
Annihilation gammas would not have collimation if the cloud annihilates
at rest and would not have the observed spectrum if the cloud is first
accelerated to a large Lorentz parameter.  The idea that the
fireball consists of electrons and positrons does, however, seem to deal with the 
absence of baryon loading so the model deserves further study.
The current situation calls for an open-minded, non-dogmatic attitude.
\end{enumerate} 
     
    Next we review
the suggestion that the strongly collimated jet structure is due to a bose
enhancement of the emitted selectrons, sprotons, and
bremstrahlung photons, i.e. a stimulated emission.  The matrix
element for the emission of a selectron pair with momenta $\vec{p}_3$ and
$\vec{p}_4$ in process
\ref{sigma} in the presence of a bath of previously emitted pairs is
proportional to

\ben
     {\cal M} &\approx& <n(\vec{p}_3)+1),n(\vec{p}_4+1)|a^\dagger(\vec{p}_3)a^\dagger(\vec{p}_4)
      |n(\vec{p}_3),n(\vec{p}_4)>\\
\nonumber\\
\nonumber\\
      &\approx& \sqrt{(n(\vec{p}_3)+1)} \sqrt{(n(\vec{p}_4)+1)}
\ee

Thus in a bath of previously emitted bosons, the 
cross section is enhanced by the factor  $(n(\vec{p}_3)+1)(n(\vec{p}_4)+1)$.
We have constructed both a primitive and a less primitive monte carlo for this
enhancement as discussed below. 

       In ref. \cite{CK}, we considered a purely statistical model where
events were generated according to this changing enhancement factor in the
three dimensional space of $\vec{p}_3$ assuming $\vec{p}_4 = -\vec{p}_3$.
One should consult ref.\cite{CK} for technical details.
This simplified statistical model has no dynamical input but
serves as a demonstration in principle of the stimulated
emission.
Initially all the $n's$ are zero but once the first transition has
been made populating a chosen $\vec{p}_3$, the next transition is
four times as likely to be into the same state as into any other
state.  Because of the huge number of available states, the second
transition is still not likely to be into the same $\vec{p}_3$
state, but as soon as some moderate number of selectrons have been
created with a common $\vec{p}_3$, the number in that state
escalates rapidly, producing a narrow jet of selectrons.  These
selectrons would be expected to decay down to the ground state via 
bremstrahlung
photons which are also Bose enhanced leading to a narrow jet of
photons which can either be reflected at the domain wall or transmitted
into the broken susy phase.

\begin{center}
\begin{tabular}{|cc|cc|cc|}
\hline
      &       &       &       &       &      \\
  p   & N & $\cos\theta$ & N& $\phi$&  N    \\
 MeV  &       &       &       &       &      \\
      &       &       &       &       &      \\
 0.02 &    52 &  -0.9 &    50 &  0.16 &    33\\
 0.07 & 99608 &  -0.7 &    60 &  0.47 &    23\\
 0.12 &    32 &  -0.5 &    34 &  0.79 &    49\\
 0.17 &    35 &  -0.3 &    71 &  1.10 &    49\\
 0.22 &    58 &  -0.1 &    45 &  1.41 &    44\\
 0.27 &    52 &   0.1 & 99598 &  1.73 &    48\\
 0.32 &    31 &   0.3 &    22 &  2.04 &    46\\
 0.37 &    49 &   0.5 &    33 &  2.36 & 99604\\
 0.42 &    30 &   0.7 &    49 &  2.67 &    65\\
 0.47 &    54 &   0.9 &    39 &  2.99 &    40\\
      &       &       &       &       &      \\
\hline
\end{tabular}
\end{center}
\begin{center}Table 1. Distribution of events generated according to the
changing number of pre-existing selectrons in each momentum
space bin.  Initially, all bins are equally likely but, after
some fluctuation has given an enhancement in one bin, that bin
is efficiently locked in for subsequent events.
\end{center}

\begin{figure}[ht]
\begin{center}
\epsfxsize= 4.5in 
\leavevmode
\epsfbox{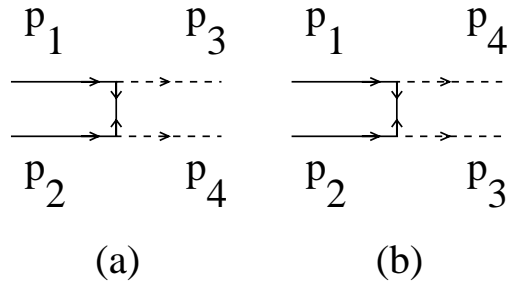}
\end{center}
\caption{$e e \rightarrow \tilde{e} \tilde{e}$ via photino exchange}
\label{Feynman}
\end{figure}

\begin{figure}[ht]
\begin{center}
\epsfxsize= 4.5in 
\leavevmode
\epsfbox{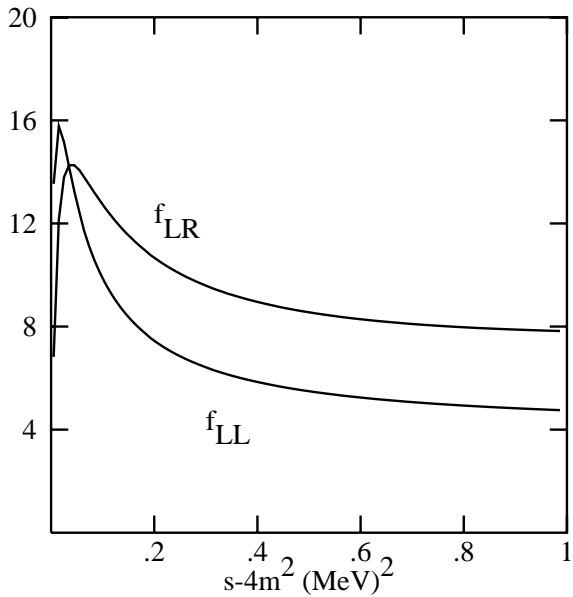}
\end{center}
\caption{helicity cross sections as a function of CM energy squared}
\label{helicity}
\end{figure}

     More physical information can be put in by calculating the electron to selectron pair conversion process in the exact susy phase where particles and sparticles have degenerate masses.
We have treated the Feynman diagrams of figure \ref{Feynman} to lowest order in the fine structure constant \cite{CP}.
The helicity cross sections near threshold are plotted in figure \ref{helicity}.

The indication that the annhilation of a left handed electron
with a right handed electron is enhanced over that of two electrons of
the same helicity suggests, in the context of stimulated emission,
a possible mechanism for a spontaneous magnetization of a white dwarf star.
A rapidly changing magnetic field could impact the long-standing mystery
of cosmic ray generation.

\begin{figure}[htb]
\begin{center}
\epsfxsize= 4.5in 
\leavevmode
\epsfbox{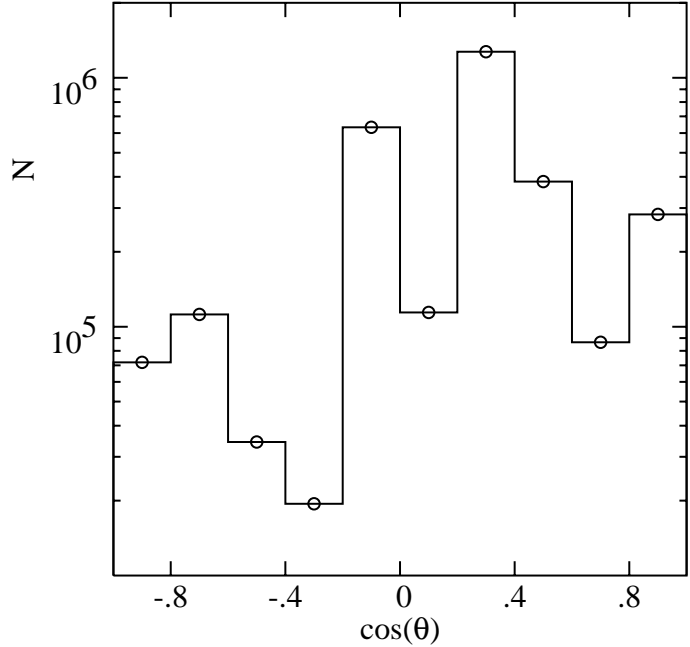}
\caption{(low resolution) angular distribution after $1.6\cdot 10^6$ events, showing 
effect of Bose enhancement}
\label{angdist}
\end{center}
\end{figure}

    We have also calculated the final state momentum distribution of the selectrons.
In this paper, the photon energy is assumed to be the
full kinetic energy of the produced selectron neglecting multiple bremstrahlung effects etc.    
Incorporating the Bose enhancement factor, the angular distribution shows
the jagged structure of figure \ref{angdist}.  Due to computer memory
limitations, the current simulations give only low resolution angular
distributions.  If there are
secondary peaks as indicated in figure \ref{angdist}, an
observer at one of the corresponding angles would interpret
the event in terms of a significantly lower ``isotropic energy''.  The latter is defined 
as the energy that would be required if
the observed burst were isotropic.  Of course, an actual
burst in the susy model would involve many more pair conversions than present in the monte-carlo of ref.\cite{CP}
and one would expect the primary peak to be further enhanced
relative to that shown in figure \ref{angdist}.  Nevertheless,
secondary peaks might explain the anomalously weak bursts
grb980425 and grb031203.   

The burst duration depends on the time taken by the susy bubble to expand to the
edge of the star and on the time taken by the gamma rays to traverse the stellar
diameter.  The minimum burst duration is, therefore, the stellar diameter divided
by the speed of light.  In actuality, of course, nothing travels at the speed of
light in dense matter.  As a mechanical membrane, the susy bubble would more reasonably
be expected to expand at the speed of sound.  Treating the star as a monatomic gas of 
constant density, the speed of sound is

\be
  v_s(r) = \sqrt{\frac{10 \pi}{9}G_N \rho (R^2-r^2)}  .
\ee

Near the center of the star the speed of sound can approach that of light; however, near
the surface the speed of sound is low.  Assuming a constant density,
the nominal white dwarf of solar mass and earth radius would have a burst duration

\be
    \tau = \int_0^{R} \frac{dr}{v_s(r)} = \frac{\pi R_E}{2 v_s(0)} \approx 2 s .
\ee

\begin{figure}[htb]
\begin{center}
\epsfxsize= 4.5in 
\leavevmode
\epsfbox{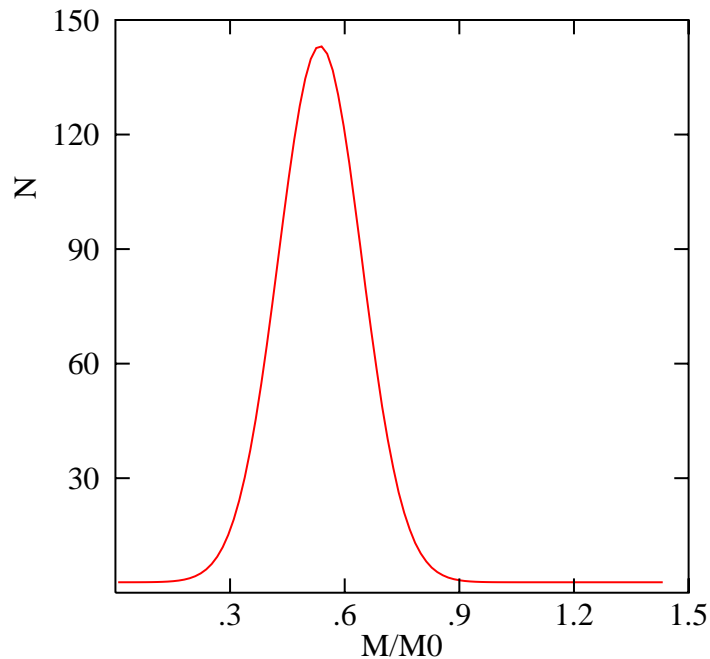}
\caption{A rough representation of the white dwarf mass distribution}
\label{massdist}
\end{center}
\end{figure}

    Up to now we have treated only the nominal white dwarf
of solar mass and earth radius.  In actuality, of course, there is a range of white dwarf
masses and radii \cite{Madej}.  The mass distribution goes from
about $0.2 M_\circ$ to nearly $1.44 M_\circ$, (the Chandrasekhar limit)
approximately as shown in figure \ref{massdist}.
It is sharply peaked at $0.56 M_\circ$, is asymmetric with some
excess on the low side, and is characterized by a long tail on both the high and low
sides.  About eighty percent of all white dwarfs lie between $0.42 M_\circ$
and $0.70 M_\circ$.  The radius of zero temperature white dwarfs ranges from about 
$0.1$ earth radius to about $2.5$ earth radii.  
The theoretical radius decreases with increasing mass.
Thus the observed white dwarfs have a wide range of densities.
Based on average densities, the burst durations would then span the range from about 
$0.2$s to $20$s.  This range coincides with the region between the
two peaks in figure \ref{duration}.  
Higher temperature white dwarfs have larger radii thus
somewhat increasing the expected burst duration range.  The dip in the duration
distribution could be affected by the interplay of the mass distribution of 
figure \ref{massdist} and the transition probability of eq. \ref{AEMB}.
This possible connection is currently under investigation.

     Since the Fermi momentum of the degenerate electron gas is
proportional to the $\frac{1}{3}$ power of the density, the
susy model allows a zeroth order understanding of the 
observation that the shorter bursts have higher average
photon energy than the long bursts.

\begin{figure}[htb]
\begin{center}
\epsfxsize= 4.5in 
\leavevmode
\epsfbox{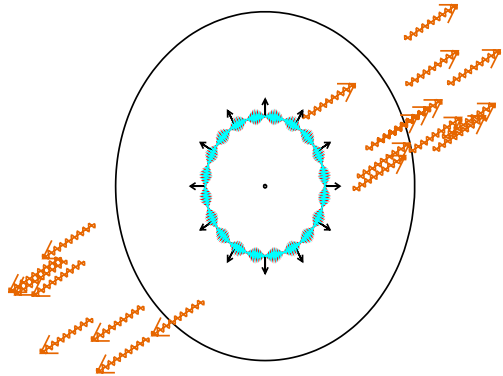}
\caption{An idealized representation of a susy phase transition in
 a dense star}
\label{grb}
\end{center}
\end{figure}

     As the next refinement, not yet completed, one must take into account 
the inhomogeneity of the white dwarfs.  The central density of such
stars is some $50$ times higher
than the average density and the density near the surface is 
correspondingly lower than average.
The susy bubble, therefore, most likely begins near the center and, most 
plausibly, expands at the speed of sound which is
rapidly deceasing as the bubble approaches the stellar
surface as indicated in the sketch of figure \ref{grb}.  The slowly expanding
bubble of true vacuum therefore constitutes a spherical cavity
inside of which photons and semi-relativistic selectrons resonate.  
Density inhomogeneity might, therefore, extend the 
burst duration distribution beyond the naive predictions given
above.  More quantitative predictions must await the
results of current studies of the effect of stellar inhomogeneity.  

      Many issues remain to be examined.  As a few examples one might
consider:
\begin{enumerate}
\item{\bf the grb-supernova connection.}  The relative rates of supernovae
$SN_{1b,c}$ to gamma ray bursts depends on the burst opening angle $\Delta
\Omega$.

\be
     \frac{R_{sn}}{R_{grb}} \approx 10^5 \frac{\Delta \Omega}{4 \pi}
\ee

If the grb collimation is
extremely narrow, these rates could be one-to-one.  
Presumably, only with a quantum collimation mechanism
such as in the susy phase transition model could such extreme
collimation be contemplated.  On the other hand, if the susy
transition takes place in a large, less dense star, the average
photon energies would be below the gamma ray range but perhaps
still more effective than neutrinos in blowing off the stellar
mantel.  In this case the ratio of supernovae to grb's could be
appreciably greater than one and one would expect a relatively
broader burst.
\item{\bf afterglows.}
  In the susy model, grb afterglows come from activated nuclei
due to the passage of the gamma ray beam through circumstellar matter
or due to fusion by-products within the star.  The former should not
be observed if there is
no appreciable circumstellar matter or if the circumstellar matter is
in a disk and the burst is perpendicular to the disk.   Since the
majority of
white dwarfs are not in binary systems, there will be many cases where there
is no afterglow of this type.  Afterglows from within the transitioning
white dwarf should not last longer than the free collapse time of a
susy star which is a function of its density.  Thus, in the susy model 
we would not expect the SWIFT 
experiment, which starts taking data soon, to see appreciable 
afterglows associated with every burst.
\end{enumerate}

    Comments on the present model at this and other conferences
have centered on ways to examine the effects of the model on other systems
and to refine the predictions for gamma ray bursts.
In addition, it is sometimes useful to consider anonymous comments since
these often give voice to critiques that their author might be ashamed to
make in an open forum.  Unfortunately, they usually have an obvious response.
Typical of these latter comments and the obvious responses to them are the following:
\begin{enumerate}

\item{``Given the large number of white dwarfs in the universe these very
common astrophysical objects are among the best studied and better
understood objects in the sky.  Their equation of state  can be tested to an
high degree of accuracy and the stars monitored for long period of time.
These observations exclude any exotic behaviour.''}

    If the author of this comment were to consult an expert, he would
immediately learn 
that the observed rate of grb's is very low compared to the 
number of white dwarfs.  This rate is

\be
     R_{grb} \approx 5 \cdot 10^{-7} \frac{0.0239}{\Delta \Omega} {\displaystyle
                 yr^{-1} galaxy^{-1}} 
\ee
with $0.0239$ being the solid angle corresponding to a $5^\circ$ burst half 
opening angle.
Even if the bursts are extremely
collimated ($\Delta \Omega \approx 4 \pi \cdot 10^{-5}$), this corresponds to
far fewer than one burst per century per galaxy.  Given the enormous number of
white dwarfs in a typical galaxy, many of which are quite dim, the probability that a given burst
could be associated with the disappearance of a known white dwarf is
negligible.  The estimates are that $95\%$ to $99\%$ of the $10^{11}$ stars 
in our galaxy are or will become white dwarfs with perhaps $10^9$ already being white dwarfs.
Until the phase transition occurs, the white dwarf star will cool according
to standard model physics.  Nevertheless, a surprising shortage of old white dwarfs has
also been observed \cite{Winget} leading to an anomalously low estimate for
the age of the universe.  It is not clear whether this is due to low
statistics or some other cause.

\item{``Supersymmetry in nature is badly broken and the splitting among the
supersymmetric partners is, at least, of the order of a TeV! So, there is no
reason to expect the particles and the sparticles to be degenerate. Besides
why are their common masses chosen, in the false vacuum, to be identical to
the ones of the particles in the ordinary vacuum and not at the TeV scale?''.}
     
     At first, the author of this comment did not seem to realize that
we are dealing with an exact susy phase where
particles and sparticles are degenerate.  Then, in
mid-sentence so to speak, he seemed to realize this basic fact but he
still did not realize that we viewed the exact susy phase as the true vacuum, not
the "false vacuum".  Finally, he did not seem to remember that superstring theory 
predicts light (even massless) ground state supermultiplets.  Ultimately, of 
course, there is an assumption in our work
that the common mass is not greater than the particle mass in the broken phase.  
Even without the string theory bias, however, this assumption is no worse than
any other and does not make our proposal unworthy of consideration.

\item{``the vacuum structure of SUSY extensions of the standard model
usually does *not* have a vacuum with unbroken supersymmetry.''}

     Even if this were ``usually'' true, it would never be used to discourage
discussion of a model except behind the mask of anonymity.
The author of this comment had in mind the highly restricted minimal
supersymmetric standard models (MSSM) with two higgs and fixed
susy breaking parameters.
Obviously, such models have no dynamical
mechanism for a phase transition to exact susy.   
In fact, of course, the five basic string
theories are supersymmetric and string models have been constructed
with standard-model-like broken supersymmetry (e.g. \cite{Nilles}
and references therein).  
In string theory all the parameters of the theory are dynamically
determined.  A formal string theory
study of the transition we consider (from an unstable
de Sitter vacuum with positive vacuum energy to a stable susy vacuum) 
has already been undertaken \cite{KKLT} and well over a hundred
articles have been written in the past two years on string landscape ideas.  
Our work differs from these only in that we suggest immediate
experimental consequences.
\end{enumerate}

   In summary, we feel that the field of
gamma ray bursts may still be in its infancy despite more than
three decades of intense study. Therefore, some 
tolerance of new physics proposals may be in order.  The mystery
is there to be solved and not just to be savoured and protected
from radical new ideas.

   This work was partially supported by the DOE under grant number
DE-FG02-96ER-40967.  We acknowledge important discussions with
George Karatheodoris, Irina Perevalova, Yongjoo Ou, Gene Byrd, 
Bill Keel, and Hans-Peter Nilles.

\end{document}